%% file: Pursuing_Equilibrium_of_Medical_Resources_via_Data_Empowerment_in_Parallel_Healthcare_System.tex
\documentclass[conference]{IEEEtran}
\IEEEoverridecommandlockouts
\usepackage{cite}
\usepackage{amsmath,amssymb,amsfonts}
\usepackage{algorithmic}
\usepackage{graphicx}
\usepackage{textcomp}
\usepackage{xcolor}

\usepackage{array}
\usepackage{booktabs}
\usepackage{tabularx}
\usepackage{multirow}

\usepackage{algorithm}
\usepackage{mathtools}
\usepackage{amsmath}

\usepackage{subcaption}
\usepackage{graphicx}

\begin{document}

\title{Pursuing Equilibrium of Medical Resources via Data Empowerment in Parallel Healthcare System}

\author{
\IEEEauthorblockN{Yi Yu}
\IEEEauthorblockA{\textit{Urban Computing Lab} \\
\textit{Shanghai AI Laboratory}\\
Shanghai, China \\
yuyi@pjlab.org.cn}
\and
\IEEEauthorblockN{Shengyue Yao\textsuperscript{\dag}}
\IEEEauthorblockA{\textit{Urban Computing Lab} \\
\textit{Shanghai AI Laboratory}\\
Shanghai, China \\
yaoshengyue@pjlab.org.cn}
\and
\IEEEauthorblockN{Kexin Wang\textsuperscript{\dag}}
\IEEEauthorblockA{\textit{School of Health Policy and Management} \\
\textit{Nanjing Medical University}\\
\textit{Nanjing, China}\\
jsxzsnwkx@163.com}
\and
\IEEEauthorblockN{4\textsuperscript{th} Yan Chen\textsuperscript{*}}
\IEEEauthorblockA{\textit{Nanjing Medical University} \\
\textit{Affiliated Suzhou Hospital, Suzhou, China}\\
\textit{School of Health Policy and Management,}\\
\textit{Nanjing Medical University, Nanjing, China}\\
chenyandoc@163.com}
\and
\IEEEauthorblockN{5\textsuperscript{th} Fei-Yue Wang \textsuperscript{*}}
\IEEEauthorblockA{\textit{SKL-MCCS, Institute of Automation} \\
\textit{Chinese Academy of Sciences, Beijing, China}\\
\textit{The Macau Institute of Systems Engineering,}\\
\textit{Macau University of Science and Technology} \\
 Macau, China, feiyue.wang@ia.ac.cn \\
}
\and
\IEEEauthorblockN{6\textsuperscript{th} Yilun Lin\textsuperscript{*}}
\IEEEauthorblockA{\textit{Urban Computing Lab} \\
\textit{Shanghai AI Laboratory}\\
Shanghai, China \\
linyilun@pjlab.org.cn}
\thanks{$\dag$ for co-first author, * for corresponding author}
}
\maketitle

\begin{abstract}
The imbalance between the supply and demand of healthcare resources is a global challenge, which is particularly severe in developing countries. Governments and academic communities have made various efforts to increase healthcare supply and improve resource allocation. However, these efforts often remain passive and inflexible. Alongside these issues, the emergence of the parallel healthcare system has the potential to solve these problems by unlocking the data value. The parallel healthcare system comprises Medicine-Oriented Operating Systems (MOOS), Medicine-Oriented Scenario Engineering (MOSE), and Medicine-Oriented Large Models (MOLMs), which could collect, circulate, and empower data. In this paper, we propose that achieving equilibrium in medical resource allocation is possible through parallel healthcare systems via data empowerment. The supply-demand relationship can be balanced in parallel healthcare systems by (1) increasing the supply provided by digital and robotic doctors in MOOS, (2) identifying individual and potential demands by proactive diagnosis and treatment in MOSE, and (3) improving supply-demand matching using large models in MOLMs. To illustrate the effectiveness of this approach, we present a case study optimizing resource allocation from the perspective of facility accessibility. Results demonstrate that the parallel healthcare system could result in up to 300\% improvement in accessibility.

\end{abstract}

\begin{IEEEkeywords}
parallel healthcare system, parallel intelligence, equilibrium, medical resource allocation, 
\end{IEEEkeywords}

\section{Introduction}
The impact of urbanization and population aging on healthcare services is becoming increasingly apparent, marked by the great gap between the supply and demand of medical resources, particularly severe in developing countries\cite{carrillo-larcoEstimatingGapDemand2022,monsefHealthcareServicesGap2023}. Firstly, the gap is caused by limited medical resource supply due to insufficient medical staff, equipment, and facilities\cite{carrillo-larcoEstimatingGapDemand2022,monsefHealthcareServicesGap2023}. For instance, over 20\% of countries lack physicians to provide hypertension patients with at least one annual medical consultation \cite{carrillo-larcoEstimatingGapDemand2022}. Additionally, the gap is exacerbated by mismatches between supply and demand, such as redundant diagnostic procedures and imbalanced medical facility layouts. Consequently, healthcare quality has declined, and doctor-patient relationships have become strained, highlighting the need for a healthcare system reform\cite{worldhealthorganizationWorldHealthReport2010}.

To alleviate the medical resource gap, governments and academic communities conduct substantial efforts from different perspectives.
Governments increase healthcare expenditure to increase the supply, including the expansion of healthcare institutions and procurement of medical equipment. For example, China issued the "Health China 2030" plan to establish a comprehensive medical service system that covers urban and rural areas \cite{HealthChina2030}. 
Academic communities conducted numerous studies focusing on optimizing the supply and allocation of medical resources, such as healthcare facility layout optimization, outpatient service procedures design, and medical diagnostic model construction\cite{huUseRealTimeInformation2023, kulkarniArtificialIntelligenceMedicine2020, zhangGainingRationalHealth2021,hongCreatingResidentShift2019,zhouResearchSpatialLayout2023}.
However, certain challenges still remain unresolved. One issue is that most optimization methods tend to be passive in nature, which are primarily designed to respond and take action only when a problem arises\cite{raghupathiBigDataAnalytics2014}. These passive approaches often fail to address potential issues proactively or utilize systematic and well-planned strategies\cite{raghupathiBigDataAnalytics2014}. Furthermore, the efforts often concentrate on optimizing specific aspects without considering the need for achieving a global equilibrium by simultaneously adjusting the supply, demand, and matching processes. These challenges underscore the need for healthcare system reforms that can enhance the fairness and efficiency of healthcare services.

Alongside these problems and needs, there are great opportunities with a surge of new intelligent technology facilities, exemplified by Artificial Intelligence(AI). Conversational AI, such as ChatGPT, offers opportunities to enhance the supply of medical resources via preprocessing information and transforming communication forms \cite{wangWhatDoesChatGPT2023}. New foundational models, such as Graph Transformers, make it possible to improve resource scheduling via automatically identifying areas for \cite{luParallelFactoriesSmart2022, wangWhatDoesChatGPT2023}. Scenario Engineering ensures the trustworthiness of AI techniques, and Cyber-Physical-Social Systems (CPSS) provide insight into transforming traditional hospital information systems into Medicine-Oriented Operating Systems\cite{liArtificialIntelligenceTest2018, liParallelTestingVehicle2019, liParallelLearningPerspective2017, liFeaturesEngineeringScenarios2022, liParallelTestingVehicle2019b, wangParallelHospitalACPBased2021,geHypertensionParallelHealthcare2022}. These burgeoning technologies support the parallel healthcare systems, which consist of 5\% biological doctors, 15\% robotic doctors, and a substantial 80\% digital doctors\cite{geHypertensionParallelHealthcare2022,wangParallelHospitalACPBased2021,wangParallelHealthcareRobotic2020,wangIntelligentSystemsTechnology2013,wangParallelEconomicsNew2020,wangParallelMedicalDiagnostic2020,wangBlockchainPoweredParallelHealthcare2018}. The parallel healthcare system will have capabilities for describe (personalized demand identification), predict (potential demand prediction), and prescribe (intelligent healthcare supply). It can offer a pathway to the medical resource equilibrium by boosting the medical resource supply, as shown in Fig. \ref{supply_demand}.  
 \begin{figure}[htbp]
    \centering
    \includegraphics[width=1\linewidth]{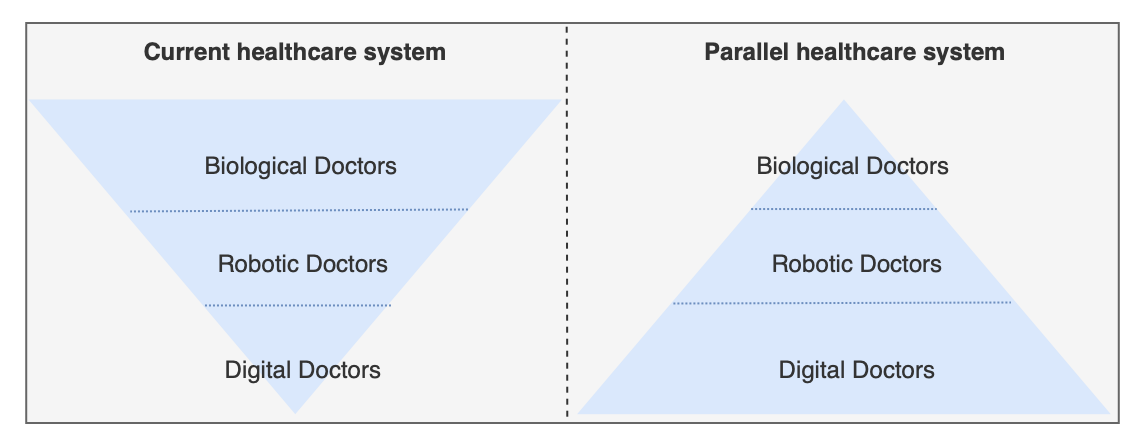}
    \caption{Differences of volume and proportion in the medical resource supply.}
    \label{change_in_supply}
\end{figure}

In this paper, we suggest that the equilibrium of medical resource allocation can be attained via the parallel healthcare system, and this optimization is essentially achieved via data empowerment. 
By collecting, circulating, and empowering large-scale medical data within parallel healthcare systems, we can unlock the value of data. The parallel healthcare systems aim to balance the supply-demand relationship through three key aspects: (1) understanding demands via describing and predicting in proactive diagnosis; (2) boosting supply via conducting treatments by digital and robotic doctors; (3) enhancing supply-demand matching using novel foundational models. By implementing parallel healthcare systems, medical resource allocation could shift from passive and fixed processes to active and asynchronous support, from symptom alleviation to health improvement. This transformation ultimately leads to a fairer and more efficient allocation of resources\cite{wangParallelHospitalACPBased2021}.

\section{Framework, Composition, and Scenarios of Parallel Healthcare System}

The essence of achieving medical resource equilibrium through parallel healthcare systems is the empowerment of data value. The parallel healthcare system encompasses Medicine-Oriented Large Models (MOLMs), Medicine-Oriented Scenario Engineering (MOSE), and Medicine-Oriented Operating Systems (MOOS). In this chapter, we will use AI technologies to incentive data circulation, utilization, and analysis in healthcare systems, thereby unlocking the value of data in optimizing medical resources.

\subsection{Framework of Parallel Healthcare System}

\begin{figure}
    \centering
    \includegraphics[width=1\linewidth]{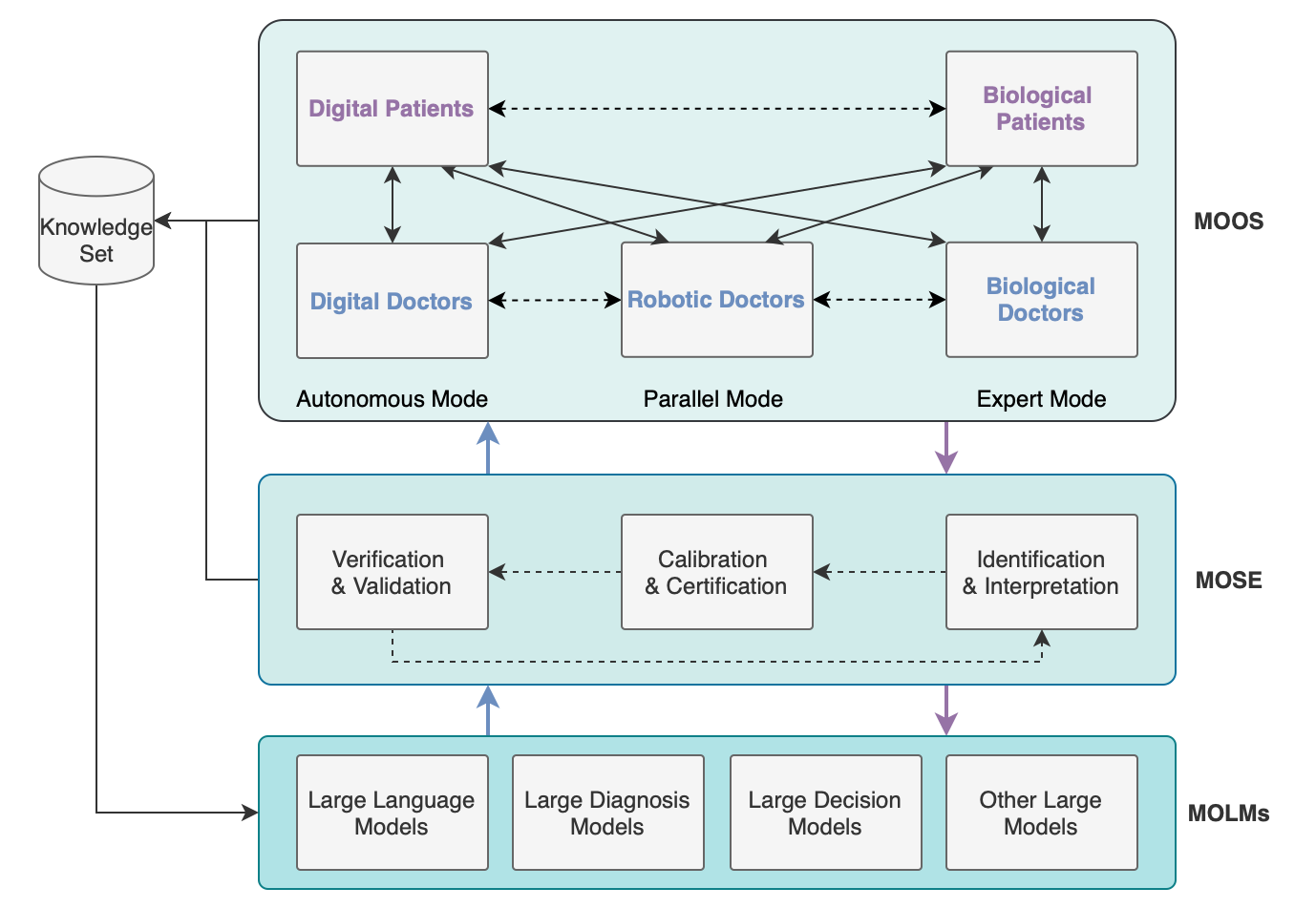}
    \caption{Parallel healthcare systems composed of Medicine-Oriented Large Models (MOLMs), Medicine-Oriented Scenario Engineering (MOSE), and Medicine-Oriented Operating Systems (MOOS). }
    \label{framework}
\end{figure}

The framework of the parallel healthcare system, as shown in Fig. \ref{framework}, reveals significant transformations in demand acquisitions and supply of medical resources, occurring in a parallel, concurrent, and asynchronous way. 
Serving as the system's gateway, MOOS acts as the backbone for efficient coordination and seamless interactions, optimizing resource allocation and decision-making processes.
MOSE plays a crucial role in decomposing complex tasks into achievable scenarios and processes, establishing robust workflows for all medical tasks. MOLMs contribute a lot by supporting the MOOS and MOSE meanwhile continuously being trained  using aggregated data records generated by MOOS and MOSE. These components of the parallel healthcare system will be elaborated upon.

\subsection{Medicine-oriented operating system}

The efficient coordination of Medical resource supply and demand requires a sophisticated infrastructure that can orchestrate various tasks and processes. At the highest level, the Medicine-Oriented Operating System (MOOS) serves as the backbone of the parallel healthcare system. MOOS organizes and optimizes the dynamic exchange of resources, facilitating seamless interactions between patients and doctors. By leveraging AI technologies, MOOS facilitates the doctor and patient virtualization, concurrent communication, and the persistence of knowledge, thus optimizing resource allocation, decision-making, and overall system efficiency.

\subsubsection{Virtualization}
The medical process in MOOS can be transformed into a virtualized form. MOOS integrates and analyzes large-scale medical data, establishing digital patient and doctor portraits. In addition, MOOS offers automatic and asynchronous approaches based on digital portraits.

\subsubsection{Concurrency}
Healthcare experts and patients can interact across time and space. On one hand, MOOS concurrently employs AI technologies to acquire demands. On the other hand, MOOS automatically, asynchronously, and concurrently provides support based on these demands.

\subsubsection{Persistence} 
MOOS digitally and persistently stores a vast amount of experts' knowledge and experience, including medical literature and case databases. Through intelligent organization and management, it supports healthcare experts in their decision-making and diagnostic processes.

The characteristics of MOOS enhance the efficiency and flexibility of healthcare, achieving the functionalities of demand acquisition and supply provision through three distinct modes: Autonomous Mode (AM), Parallel Mode (PM), and Expert Mode (EM). In Autonomous Mode, MOOS facilitates efficient and autonomous resource allocation by autonomously acquiring medical demands and providing healthcare supplies based on predefined goals and constraints. In Parallel Mode, MOOS has the capability to enhance decision-making and improve the quality of diagnosis and treatment by employing a collaborative approach where artificial intelligence collaborates with human doctors. Moreover, Expert Mode empowers experts to analyze complex medical demands and refine resource allocation through the integration of innovative algorithms. As a vital component of parallel healthcare systems, the development of MOOS by combining AM, PM, and EM modes lays the foundation for a more efficient and sustainable allocation of medical resources.

\subsection{Medicine-oriented scenario engineering}

To effectively address the diverse demands and supply of medical resources, Medicine-Oriented Scenario Engineering (MOSE) plays a crucial role\cite{liNovelScenariosEngineering2023}. MOSE decomposes complex tasks into achievable scenarios and processes, ensuring compatibility with predefined criteria. By carefully designing and validating these scenarios, MOSE establishes a robust framework for resource allocation and optimization\cite{liFeaturesEngineeringScenarios2022,changMetaScenarioFrameworkDriving2023,liNovelScenariosEngineering2023}. 

\subsubsection{Identification and Interpretation (I\&I)} MOOS could identify and interpret scenarios from patients' preferences, historical, and real-time data that align with individual demands and system requirements.

\subsubsection{Calibration and Certification (C\&C)}
MOSE could design and develop algorithms and models that can calibrate and certify according to abstract scenarios, catering to diverse medical settings. It enables efficient allocations of healthcare resources.

\subsubsection{Verification and Validation (V\&V)} 
To validate the capabilities and feasibility, MOSE employs real or simulated medical data and scenarios for experiments. The outcomes derived from Verification and Validation can be utilized for correction, guaranteeing the system's accuracy of healthcare resource allocation challenges in practical applications.

\subsection{Medicine-Oriented Large Models}
Underlying the MOOS and MOSE framework are Medicine-Oriented Large Models (MOLMs). MOLMs encompass comprehensive representations of the medical system, incorporating vast amounts of data, behavioral patterns, and domain knowledge. Using language models, diagnostic models, and decision models as examples, these large-scale models are trained using aggregated data records generated by MOOS and MOSE. Through continuous learning and refinement, MOLMs contribute to adaptability and responsiveness.

\subsubsection{Large language models}
\begin{figure}
    \centering
    \includegraphics[width=0.7\linewidth]{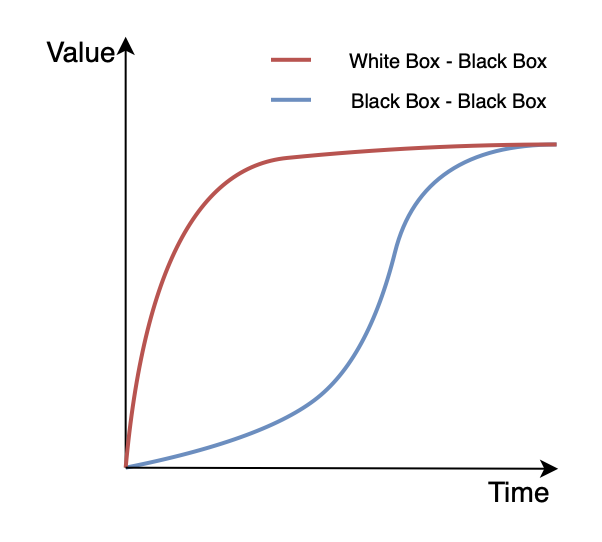}
    \caption{Cumulative value functions in conversations: In human-human interactions, communication involves incremental information disclosure, which requires a substantial amount of low-value information. In human-machine (white box - black box) interactions, communication can directly address value information based on records.}
    \label{value_function}
\end{figure}

Language Large Models (LLMs) can capture the intricate dynamics of resource supply and demand through natural language processing and machine learning techniques\cite{wangChatGPTComputationalSocial2023,wangWhatDoesChatGPT2023}. By doing so, it transforms the patient-doctor communication from a 'black box (Human)-black box' interaction to a 'white box (LLMs)-black box' interaction\cite{rajprabhuDefiningQuantifyingConversation2021}. 
LLMs can also integrate multiple data sources,  characterized by a high initial slope of the cumulative value function. In the new communication form, LLMs could leverage history data, provide real-time responsiveness, and integrate multiple data sources from other MOLMs. It results in highly efficient communication from the very beginning, ultimately enhancing the efficiency of healthcare resource allocation, as shown in Fig. \ref{value_function}.

\subsubsection{Large diagnosis models}
Large diagnosis models facilitate the supplementation of supply and the clarification of demand, thereby breaking down the barriers between the field of healthcare and other domains through the integration of diverse knowledge. Large diagnosis models can analyze and identify patterns and correlations within the data and offer diagnostic support.

\subsubsection{Large optimization models}
Large optimization models have the potential to enhance resource allocation by leveraging machine learning, deep learning, and related techniques for handling and analyzing the supply and demand of healthcare resources\cite{liCooperativeMultiAgentReinforcement2019}. For instance, models such as Graph Transformers and pricing mechanisms can automatically optimize resource scheduling patterns\cite{yuSWDPMSocialWelfareOptimized2023}.

\subsection{Resource allocation scenarios}
In practical applications, resource allocation can be categorized into three types: macro, meso, and micro, based on the corresponding relationship between supply and demand, as well as optimization objectives. Constructing optimization models according to these categories enables effective resolution of specific resource allocation optimization problems.

Micro resource allocation problems often involve situations where demand exceeds supply, such as hospital bed allocation, management of medication, and medical equipment allocation.

Meso resource allocation problems typically involve situations where there are multiple demands and multiple supplies. Examples include operating room scheduling, doctor's outpatient scheduling, as well as optimization of outpatient processes.

Macro resource allocation problems usually involve multi-objective optimization with multiple demands and multiple supplies, such as city-level allocation of medical resources and optimization of hospital locations. 

\begin{figure}
    \centering
    \includegraphics[width=1\linewidth]{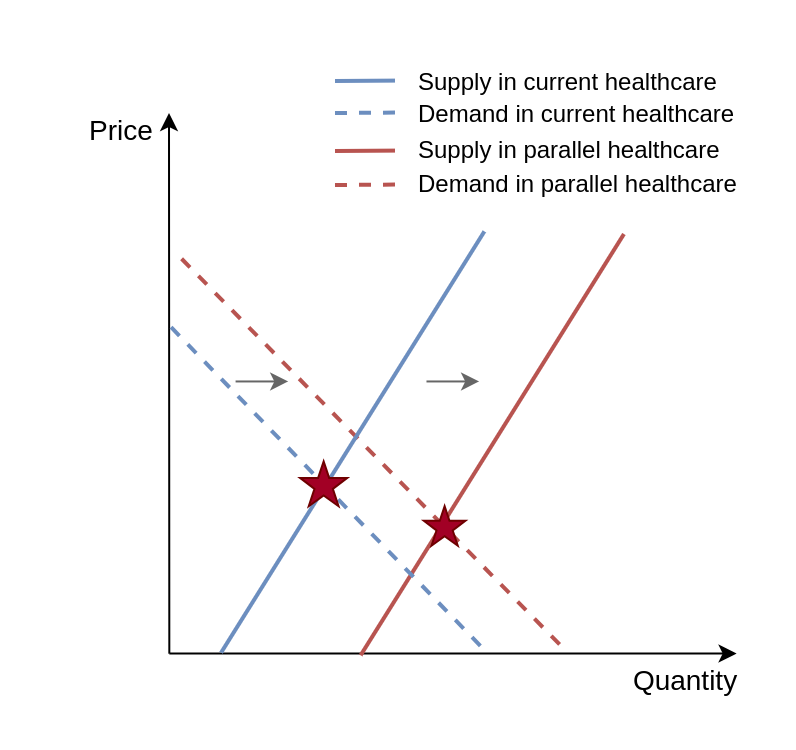}
    \caption{Parallel healthcare systems could support high-quantity and low-price services due to the shifted demand-supply curves of medical resources. (1) The supply is increased in parallel healthcare systems due to the additional treatment provided by robotic doctors and digital doctors. (2) The demand is changed because parallel healthcare systems could describe and predict potential demand, which could increase or decrease the demand value}
    \label{supply_demand}
\end{figure}

By integrating MOOS, MOSE, and MOLMs, parallel healthcare systems can collect, process, and analyze large-scale medical data. Through the application of AI technologies, the value of data in medical resources can be unlocked. The supply-demand curve of medical resources will undergo changes as shown in Fig. \ref{supply_demand}. 
The shifted demand-supply curves are caused by (1) the increasing supply in parallel healthcare systems provided by robotic doctors and digital doctors, and (2) changed demand with accurate description and prediction of potential demand, which could increase or decrease the demand. Apart from supply and demand, there have been significant transformations in the supply-demand matching of medical resources. These transformations reflect in several aspects: (1) a shift from passive to proactive medical support, (2) a departure from fixed linear processes to asynchronous concurrent support, and (3) a transition of medical knowledge from transient to persistent, which pave ways to achieve equilibrium and balance in the supply-demand relationship of medical resources.

\section{Application of Parallel Healthcare System for Medical Resource Allocation}
In this paper, we present a case study about accessibility-oriented optimization of healthcare facilities in the parallel healthcare system, as shown in Fig. \ref{framework}. The parallel healthcare system could ensure the expansion and optimization of healthcare facilities, supported by digital doctors and robotic doctors. 

\begin{figure}
    \centering
    \includegraphics[width=1\linewidth]{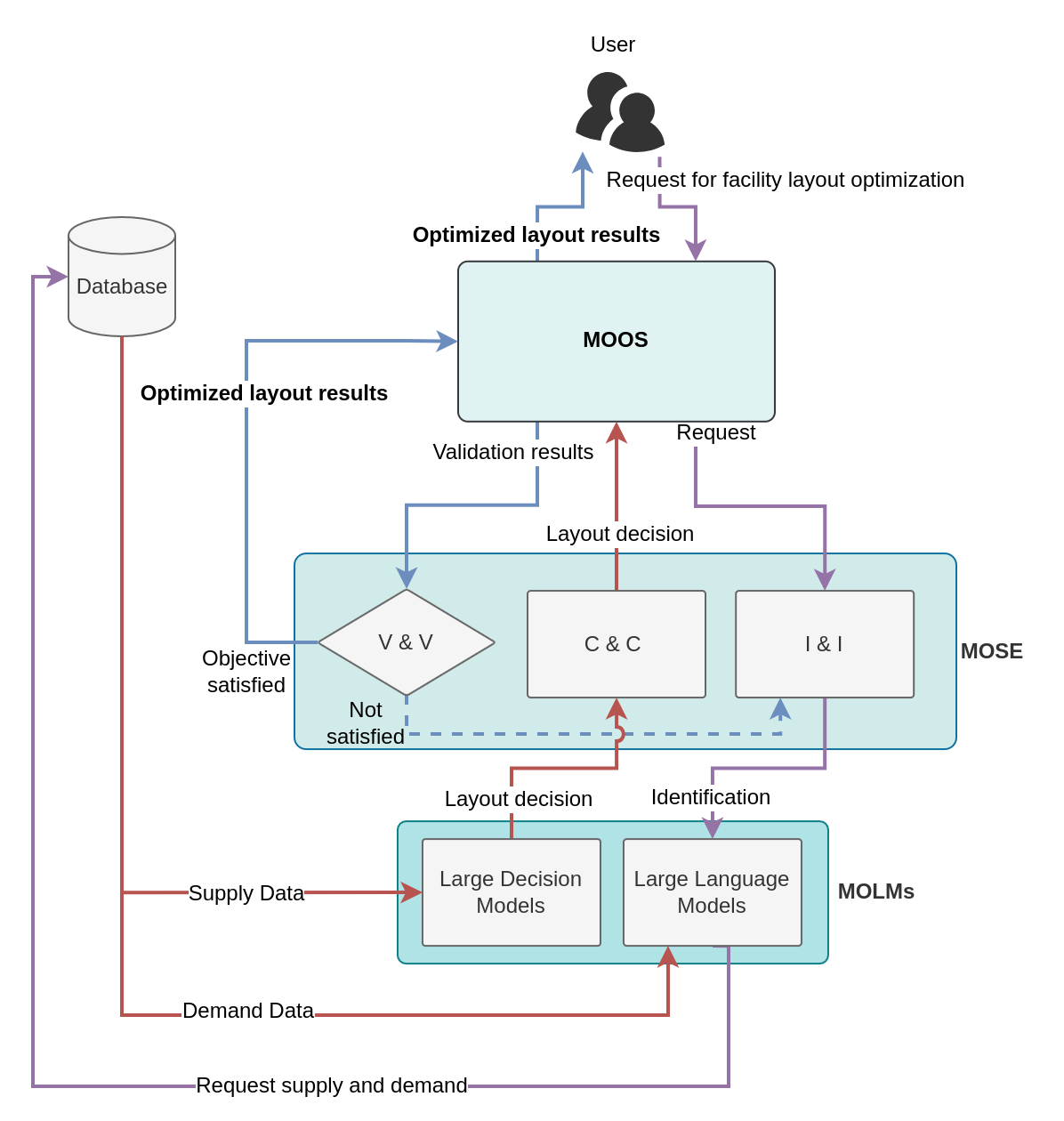}
    \caption{Application workflow of parallel healthcare systems for optimizing medical resource allocation}
    \label{bar}
\end{figure}

\subsection{Accessibility definition and objectives}
The accessibility of healthcare resources can be obtained by computing the attributes of healthcare resource demand and supply in spatial dimensions, using the Two-step Floating Catchment Area (2SFCA) method. 

The accessibility $A_{i}$ based on the 2SFCA method is formulated as:
\begin{subequations}
\label{accesibility_function}
\begin{equation}
    A_{i}=\sum_{j\in {t{\leq }t_\sigma}}\frac{ \gamma G(t_{ij}) S_{j}}{\sum_{n\in \{t_{nj}\leq t_\sigma\}}^{}G(t_{ij})D_{n}}
\end{equation}
where
\begin{equation}
     G(t_{ij})=\frac{e^{-\frac{1}{2}\times \left ( \frac{t_{ij}}{t_\sigma}\right )^{2}}-e^{-\frac{1}{2}}{}}{1-e^{-\frac{1}{2}}}
\end{equation}
\end{subequations}
where $G(t_{ij})$ is the time-decay function,  quantifying the decay in accessibility as the travel time $t_{ij}$ between locations $i$ and $j$ increases;  $S_{j}$ is the number of healthcare facilities at location $j$; $\gamma$ is the parameter that adjusts the weight of the supply-demand ratio; $t_\sigma$ is a travel time threshold to determine whether locations $j$ should be considered; $n$ is the index representing the district is in the area where its travel time $t_{nj}\leq t_\sigma$; $D_n$ is the volume of demand represented by population within the catchment area of location.

To achieve a balance between accessibility and fairness in healthcare facilities, the objective function is set as\ref{objective_function}:
\begin{equation}
\begin{aligned}
\min \ &\alpha \cdot k + \beta \cdot \sum_{i=1}^{m} (\Delta A_i)^2 \\
\text{s.t.} \quad &\forall i \in m \quad A_i \geq A_\sigma
\end{aligned}
\label{objective_function}
\end{equation}
where $\alpha$ and $\beta$ are weights in the objective function; $m$ is the number of demand locations; $\Delta A_i= |A_i - A_\sigma|$ in which $A_{\sigma}$ is the target accessibility and $A_i$ is the accessibility in location $i$.

\subsection{Experiment Setting}
We conduct the case study in the Gulou District, Nanjing. The district spans an area of 54.18 $km^{2}$ and accommodates a population exceeding 900,000 individuals. We take the population as a representative of the medical resource demand, and the data on population distribution were obtained on April 4, 2023. Previous COVID test huts are regarded as available medical resource supplies. According to a survey conducted on March 9, 2023, there are currently 16 healthcare facilities in place. These huts can be repurposed as healthcare facilities to enhance the supply of medical resources.

In this study, we set target accessibility $A_\sigma = 0.135$, weights in objective function $\alpha$, and $\beta$ as 1. The capability of the healthcare facility of supplying is set as 1500 individuals per day. 
In addition, we determine that the average walking speed for the general public is 80m per minute, with a maximum walking distance of 700m \cite{lu2012walkability}. For the elderly, the average walking speed is 70m per minute, with a maximum walking distance also limited to 700m \cite{lu2012walkability}. By utilizing the road network data from OpenStreetMap the accessibility can be computed.

\subsection{Evaluation results}
Based on the aforementioned method and objective function, in the parallel healthcare system, we could  achieve both accessibility and fairness of healthcare facilities layout by only constructing 15 additional healthcare facilities, which means a total of 31 healthcare facilities. 

The parallel healthcare system further gives out an optimized facility layout. Comparisons of accessibility results are demonstrated in Fig. \ref{bar}, revealing an overall optimization of accessibility. Before optimizations, only 3.65\% of residents have a very high level of accessibility, indicating a deficiency in fairness, as indicated by Fig. \ref{bar}. Following the establishment of 15 healthcare facilities using parallel healthcare systems, the overall coverage for the general public notably increased to over 95\%. The range of highly accessible areas expanded significantly, with up to a 300\% increase in coverage for populations above medium accessibility. 

Regarding the elderly population, their overall accessibility level was considerably lower before optimization. The coverage was limited, with only approximately 7.27\% of elderly individuals able to reach healthcare facilities within 10 minutes, as illustrated in Fig. \ref{bar}. The prevalence of low accessibility was notably 70.47\%, revealing imbalanced resource allocation. After optimization, there was a significant improvement with overall coverage increased to 95\% or higher. 

\begin{figure}
    \centering
    \includegraphics[width=0.8\linewidth]{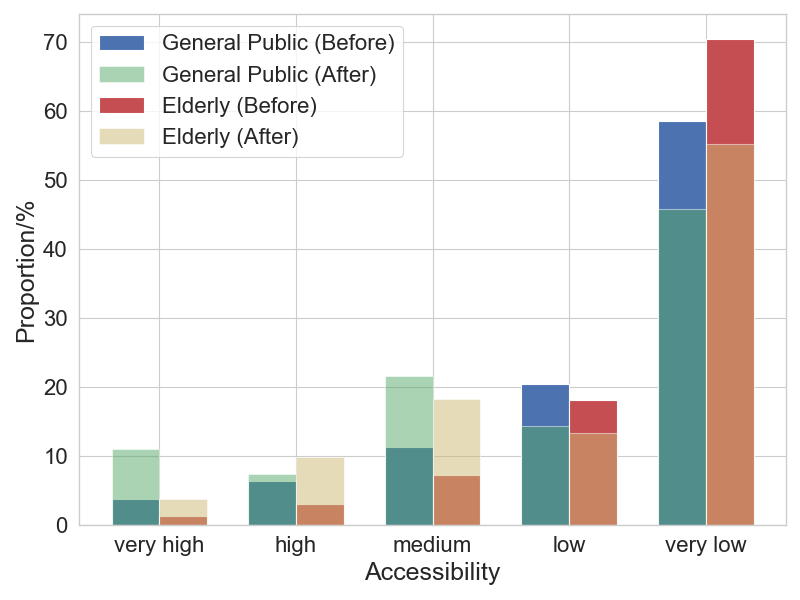}
    \caption{Population proportions of different accessibility coverage levels for healthcare facilities}
    \label{bar}
\end{figure}

In summary, significant improvements have been achieved in the overall accessibility for both the general public and the elderly.
These improvements address parallel healthcare systems' potential of approaching medical resources equilibrium.

\section{Conclusion and future works}
In this paper, we discuss the parallel healthcare system, which presents a promising path forward equilibrium in medical resource allocation via data empowerment. By analyzing the supply and demand of medical resources and leveraging AI technologies, we proposed a parallel healthcare system framework to balance the supply-demand relationship. 

Through the implementation of parallel healthcare systems, which include Medicine-Oriented Operating Systems (MOOS), Medicine-Oriented Scenario Engineering (MOSE), and Medicine-Oriented Large Models (MOLMs), we can boost resource supply, optimize allocation, predict demand, and improve medical resource allocation, leading to a more efficient and social-centric healthcare system. An application of medical facilities optimization demonstrated a threefold improvement in accessibility, showcasing the potential of parallel healthcare systems to optimize medical resource allocation.

While parallel healthcare systems offer a promising approach to optimizing medical resource allocation, there is still much work left for the future. (1) Constructing MOOS that aligns with practical application requirements (2) Integrating MOSE with domain knowledge in medicine to facilitate smoother workflows. (3) Exploring solutions to privacy-preserving problems when using large models in MOLMs. We expect the community to delve deeper into this field, extending and refining its framework and applications. Through continuous exploration and innovation in parallel healthcare systems, we can unlock its full potential and establish a more equitable and efficient healthcare system worldwide.

\section*{Acknowledgment}
This work is supported by the Shanghai Artificial Intelligence Laboratory, 2022 Decision Consultation and Cultivation Project of Jiangsu Provincial Institute of Health(JX102C20160020), 2021 Graduate Quality Education Resource Construction Project(2021B016), Key Project of Hospital Management Innovation Research of Jiangsu Hospital Association(JSYGY-2-2020-695), and Jiangsu Graduate Research and Innovation Program(KYCX22\_1769).


\input{reference.bbl}

\bibliographystyle{IEEEtran}
\bibliography{reference}

\end{document}

%% file: reference.bbl

%% file: Pursuing_Equilibrium_of_Medical_Resources_via_Data_Empowerment_in_Parallel_Healthcare_System.bbl
\begin{thebibliography}{10}
\providecommand{\url}[1]{#1}
\csname url@samestyle\endcsname
\providecommand{\newblock}{\relax}
\providecommand{\bibinfo}[2]{#2}
\providecommand{\BIBentrySTDinterwordspacing}{\spaceskip=0pt\relax}
\providecommand{\BIBentryALTinterwordstretchfactor}{4}
\providecommand{\BIBentryALTinterwordspacing}{\spaceskip=\fontdimen2\font plus
\BIBentryALTinterwordstretchfactor\fontdimen3\font minus
  \fontdimen4\font\relax}
\providecommand{\BIBforeignlanguage}[2]{{%
\expandafter\ifx\csname l@#1\endcsname\relax
\typeout{** WARNING: IEEEtran.bst: No hyphenation pattern has been}%
\typeout{** loaded for the language `#1'. Using the pattern for}%
\typeout{** the default language instead.}%
\else
\language=\csname l@#1\endcsname
\fi
#2}}
\providecommand{\BIBdecl}{\relax}
\BIBdecl

\bibitem{carrillo-larcoEstimatingGapDemand2022}
R.~M. {Carrillo-Larco}, W.~C. {Guzman-Vilca}, and D.~Neupane, ``Estimating the
  gap between demand and supply of medical appointments by physicians for
  hypertension care: A pooled analysis in 191 countries,'' \emph{BMJ Open},
  vol.~12, no.~4, p. e059933, Apr. 2022.

\bibitem{monsefHealthcareServicesGap2023}
N.~Monsef, E.~Suliman, E.~Ashkar, and H.~Y. Hussain, ``Healthcare services gap
  analysis: A supply capture and demand forecast modelling, {{Dubai}}
  2018\textendash 2030,'' \emph{BMC Health Services Research}, vol.~23, no.~1,
  p. 468, May 2023.

\bibitem{worldhealthorganizationWorldHealthReport2010}
{World Health Organization}, ``The world health report: health systems
  financing: the path to universal coverage,'' World Health Organization, Tech.
  Rep., 2010.

\bibitem{HealthChina2030}
``Health china 2030,''
  http://www.gov.cn/zhengce/2016-10/25/content\_5124174.htm.

\bibitem{huUseRealTimeInformation2023}
Y.~Hu, K.~D. Cato, C.~W. Chan, J.~Dong, N.~Gavin, S.~C. Rossetti, and B.~P.
  Chang, ``Use of real-time information to predict future arrivals in the
  emergency department,'' \emph{Annals of Emergency Medicine}, pp.
  S0196--0644(22)01\,269--0, Jan. 2023.

\bibitem{kulkarniArtificialIntelligenceMedicine2020}
S.~Kulkarni, N.~Seneviratne, M.~S. Baig, and A.~H.~A. Khan, ``Artificial
  intelligence in medicine: Where are we now?'' \emph{Academic Radiology},
  vol.~27, no.~1, pp. 62--70, Jan. 2020.

\bibitem{zhangGainingRationalHealth2021}
Y.~Zhang, H.~Yang, and J.~Pan, ``Gaining from rational health planning: Spatial
  reallocation of top-tier general hospital beds in china,'' \emph{Computers \&
  Industrial Engineering}, vol. 157, p. 107344, Jul. 2021.

\bibitem{hongCreatingResidentShift2019}
Y.-C. Hong, A.~Cohn, M.~A. Epelman, and A.~Alpert, ``Creating resident shift
  schedules under multiple objectives by generating and evaluating the pareto
  frontier,'' \emph{Operations Research for Health Care}, vol.~23, p. 100170,
  Dec. 2019.

\bibitem{zhouResearchSpatialLayout2023}
Q.~Zhou and Y.~Zheng, ``Research on the spatial layout optimization strategy of
  huaihe road commercial block in hefei city based on space syntax theory,''
  \emph{Frontiers in Computational Neuroscience}, vol.~16, 2023.

\bibitem{raghupathiBigDataAnalytics2014}
W.~Raghupathi and V.~Raghupathi, ``Big data analytics in healthcare: Promise
  and potential,'' \emph{Health Information Science and Systems}, vol.~2,
  no.~1, p.~3, Feb. 2014.

\bibitem{wangWhatDoesChatGPT2023}
F.-Y. Wang, Q.~Miao, X.~Li, X.~Wang, and Y.~Lin, ``What does chatgpt say: The
  dao from algorithmic intelligence to linguistic intelligence,''
  \emph{IEEE/CAA Journal of Automatica Sinica}, vol.~10, no.~3, pp. 575--579,
  Mar. 2023.

\bibitem{luParallelFactoriesSmart2022}
J.~Lu, X.~Wang, X.~Cheng, J.~Yang, O.~Kwan, and X.~Wang, ``Parallel factories
  for smart industrial operations: From big ai models to field foundational
  models and scenarios engineering,'' \emph{IEEE/CAA Journal of Automatica
  Sinica}, vol.~9, no.~12, pp. 2079--2086, Dec. 2022.

\bibitem{liArtificialIntelligenceTest2018}
L.~Li, Y.-L. Lin, N.-N. Zheng, F.-Y. Wang, Y.~Liu, D.~Cao, K.~Wang, and W.-L.
  Huang, ``Artificial intelligence test: A case study of intelligent
  vehicles,'' \emph{Artificial Intelligence Review}, vol.~50, no.~3, pp.
  441--465, 2018.

\bibitem{liParallelTestingVehicle2019}
L.~Li, X.~Wang, K.~Wang, Y.~Lin, J.~Xin, L.~Chen, L.~Xu, B.~Tian, Y.~Ai, and
  J.~Wang, ``Parallel testing of vehicle intelligence via virtual-real
  interaction,'' \emph{Science robotics}, 2019.

\bibitem{liParallelLearningPerspective2017}
L.~Li, Y.~Lin, N.~Zheng, and F.-Y. Wang, ``Parallel learning: A perspective and
  a framework,'' \emph{IEEE/CAA Journal of Automatica Sinica}, vol.~4, no.~3,
  pp. 389--395, 2017.

\bibitem{liFeaturesEngineeringScenarios2022}
X.~Li, P.~Ye, J.~Li, Z.~Liu, L.~Cao, and F.-Y. Wang, ``From features
  engineering to scenarios engineering for trustworthy ai: I\&i, c\&c,; v\&v,''
  \emph{IEEE Intelligent Systems}, vol.~37, no.~4, pp. 18--26, Jul. 2022.

\bibitem{liParallelTestingVehicle2019b}
L.~Li, X.~Wang, K.~Wang, Y.~Lin, J.~Xin, L.~Chen, L.~Xu, B.~Tian, Y.~Ai,
  J.~Wang, D.~Cao, Y.~Liu, C.~Wang, N.~Zheng, and F.-Y. Wang, ``Parallel
  testing of vehicle intelligence via virtual-real interaction,'' \emph{Science
  Robotics}, vol.~4, no.~28, p. eaaw4106, Mar. 2019.

\bibitem{wangParallelHospitalACPBased2021}
Y.~Wang, F.-Y. Wang, X.~Wang, Z.~Guan, L.~Ouyang, J.~Wang, L.~Yan, W.~Zheng,
  and W.~Zhang, ``Parallel hospital: Acp-based hospital smart operating
  system,'' in \emph{2021 IEEE 1st International Conference on Digital Twins
  and Parallel Intelligence (DTPI)}, Jul. 2021, pp. 474--477.

\bibitem{geHypertensionParallelHealthcare2022}
L.~Ge, B.~Lv, N.~Li, S.~An, and F.-Y. Wang, ``A hypertension parallel
  healthcare system based on the acp approach,'' \emph{IEEE Journal of Radio
  Frequency Identification}, vol.~6, pp. 724--728, 2022.

\bibitem{wangParallelHealthcareRobotic2020}
F.-Y. Wang, ``Parallel healthcare: Robotic medical and health process
  automation for secured and smart social healthcares,'' \emph{IEEE
  Transactions on Computational Social Systems}, vol.~7, no.~3, pp. 581--586,
  Jun. 2020.

\bibitem{wangIntelligentSystemsTechnology2013}
F.-Y. Wang and P.~K. Wong, ``Intelligent systems and technology for integrative
  and predictive medicine: An acp approach,'' \emph{ACM Transactions on
  Intelligent Systems and Technology}, vol.~4, no.~2, pp. 32:1--32:6, Apr.
  2013.

\bibitem{wangParallelEconomicsNew2020}
F.-Y. Wang, ``Parallel economics: A new supply\textendash demand philosophy via
  parallel organizations and parallel management,'' \emph{IEEE Transactions on
  Computational Social Systems}, vol.~7, no.~4, pp. 840--848, Aug. 2020.

\bibitem{wangParallelMedicalDiagnostic2020}
J.~Wang, X.~Wang, Y.~Guo, and F.-Y. Wang, ``A parallel medical diagnostic and
  treatment system for chronic diseases,'' in \emph{2020 Chinese Automation
  Congress (CAC)}, Nov. 2020, pp. 7412--7416.

\bibitem{wangBlockchainPoweredParallelHealthcare2018}
S.~Wang, J.~Wang, X.~Wang, T.~Qiu, Y.~Yuan, L.~Ouyang, Y.~Guo, and F.-Y. Wang,
  ``Blockchain-powered parallel healthcare systems based on the acp approach,''
  \emph{IEEE Transactions on Computational Social Systems}, vol.~5, no.~4, pp.
  942--950, Dec. 2018.

\bibitem{liNovelScenariosEngineering2023}
X.~Li, Y.~Tian, P.~Ye, H.~Duan, and F.-Y. Wang, ``A novel scenarios engineering
  methodology for foundation models in metaverse,'' \emph{IEEE Transactions on
  Systems, Man, and Cybernetics: Systems}, vol.~53, no.~4, pp. 2148--2159, Apr.
  2023.

\bibitem{changMetaScenarioFrameworkDriving2023}
C.~Chang, D.~Cao, L.~Chen, K.~Su, K.~Su, Y.~Su, F.-Y. Wang, J.~Wang, P.~Wang,
  J.~Wei, G.~Wu, X.~Wu, H.~Xu, N.~Zheng, and L.~Li, ``Metascenario: A framework
  for driving scenario data description, storage and indexing,'' \emph{IEEE
  Transactions on Intelligent Vehicles}, vol.~8, no.~2, pp. 1156--1175, Feb.
  2023.

\bibitem{wangChatGPTComputationalSocial2023}
F.-Y. Wang, J.~Li, R.~Qin, J.~Zhu, H.~Mo, and B.~Hu, ``Chatgpt for
  computational social systems: From conversational applications to
  human-oriented operating systems,'' \emph{IEEE Transactions on Computational
  Social Systems}, vol.~10, no.~2, pp. 414--425, Apr. 2023.

\bibitem{rajprabhuDefiningQuantifyingConversation2021}
N.~Raj~Prabhu, C.~Raman, and H.~Hung, ``Defining and quantifying conversation
  quality in spontaneous interactions,'' in \emph{Companion Publication of the
  2020 International Conference on Multimodal Interaction}, ser. ICMI '20
  Companion.\hskip 1em plus 0.5em minus 0.4em\relax New York, NY, USA:
  Association for Computing Machinery, Dec. 2021, pp. 196--205.

\bibitem{liCooperativeMultiAgentReinforcement2019}
X.~Li, J.~Zhang, J.~Bian, Y.~Tong, and T.-Y. Liu, ``A cooperative multi-agent
  reinforcement learning framework for resource balancing in complex logistics
  network,'' Mar. 2019.

\bibitem{yuSWDPMSocialWelfareOptimized2023}
Y.~Yu, S.~Yao, J.~Li, F.-Y. Wang, and Y.~Lin, ``{SWDPM} : A social
  welfare-optimized data pricing mechanism,'' May 2023.

\bibitem{lu2012walkability}
Y.~Lu and D.~Wang, ``Walkability measuring in america and its enlightenment,''
  \emph{Urban Plan. Int}, vol.~27, pp. 10--15, 2012.

\end{thebibliography}
